\documentclass[preprint]{elsarticle}
\makeatletter
\def\ps@pprintTitle{%
 \let\@oddhead\@empty
 \let\@evenhead\@empty
 \def\@oddfoot{\centerline{\thepage}}%
 \let\@evenfoot\@oddfoot}
\makeatother
\usepackage{amsmath}
\usepackage{graphicx}
\usepackage{xfrac}
\usepackage[hidelinks]{hyperref}
\biboptions{sort&compress}
\begin{document}

\title{GPU-accelerated micromagnetic simulations using cloud computing}
\author[cu]{C. L. Jermain}
\cortext[cor1]{Corresponding author}
\ead{clj72@cornell.edu}
\author[cu]{G. E. Rowlands}
\author[cu]{R. A. Buhrman}
\author[cu,kavli]{D. C. Ralph}
\address[cu]{Cornell University, Ithaca, New York 14853, USA}
\address[kavli]{Kavli Institute at Cornell, Ithaca, New York 14853, USA}

\date{\today}

\begin{abstract}
Highly-parallel graphics processing units (GPUs) can improve the speed of micromagnetic simulations significantly as compared to conventional computing using central processing units (CPUs). We present a strategy for performing GPU-accelerated micromagnetic simulations by utilizing cost-effective GPU access offered by cloud computing services with an open-source Python-based program for running the MuMax3 micromagnetics code remotely. We analyze the scaling and cost benefits of using cloud computing for micromagnetics.
\end{abstract}


\maketitle

\section{Introduction}

Micromagnetic simulations provide quantitative predictions for complex magnetic physics\cite{Fidler2000}, including the influences of demagnetization, spin-transfer torque\cite{Berkov2008}, and the Dzyaloshinskii-Moriya interaction\cite{Thiaville2012,Perez2014}. Recent advances in graphics processing units (GPUs) have prompted the integration of such computing capacity into micromagnetic packages\cite{Vansteenkiste2013,Vansteenkiste2014,Vansteenkiste2011,Chang2011,MicroMagnum,Lopez-Diaz2012}. The massively-parallel character of GPUs is particularly well-suited to accelerating large finite-difference calculations, such as the simulation of magnetization dynamics in extended films and the full layer structures of magnetic tunnel junctions. However, GPU-based computing requires specialized hardware. Furthermore, current GPU-based simulators are based on the CUDA software library, which is restricted to NVIDIA-manufactured hardware, further limiting their accessibility. Here we discuss an approach we have developed for running micromagnetic simulations on cloud computing services, thereby eliminating the need to purchase and maintain dedicated GPU hardware. We present open-source software that allows researchers who are unfamiliar with GPUs and cloud computing to readily perform cost-efficient micromagnetic simulations from any computer. Finally, we analyze the conditions under which the GPU-based approach confers an advantage over CPU-based simulations.

While a number of GPU-accelerated micromagnetic packages have been developed, we focus on the open-source project MuMax3\cite{Vansteenkiste2011,Vansteenkiste2014}. Open-source codes are free of the licensing restrictions of commerical packages that often preclude their execution in cluster environments or on cloud computing platforms. Moreover, the availability of the source code to the scientific community allows the underlying mechanics of the simulations to be scrutinized if doubts are raised about the results.

\section{Cloud computing services}

Cloud computing services are a comprehensive set of tools for performing computations on hardware resources that are offered over the Internet\cite{Vliet2011}. Providers sell access to virtual computers, known as instances, that run on their hardware and can be launched on an on-demand or reservation basis. Instances come in a variety of hardware configurations that are reflected in their hourly prices. When using cloud computing services to perform a GPU-based micromagnetic simulation, a user first launches a GPU instance on the provider's servers. Rather than having to install the necessary software after every launch, instances can instead be based on previously created ``images'' that have the micromagnetic and supporting packages pre-installed. Simulation input files are transferred to the running instance, and the simulation runs on the remote hardware until completion. The data is then transferred back to the user's local computer. At this point the instance can be stopped to avoid incurring further hourly charges, or kept open to continue with other simulations.

Here we investigate the use of the GPU instance type offered by Amazon Web Services (AWS)\cite{AWS}. At the time of this writing, this instance type has an NVIDIA GRID K520 GPU card, with 1536 CUDA cores and 4 GB of video memory for an hourly price of \$0.65. A consumer NVIDIA card with comparable specifications, the NVIDIA GTX 770, retails for \$310 $\pm$ 20. At the lower bound, where only the GPU cost is considered, a researcher would need to perform over 480$\pm$30 hours of simulations on AWS before the cost of the graphics card is recovered. This calculation considers neither the cost of a desktop computer able to house the GPU nor the ongoing maintenance costs thereof. AWS has upgraded their GPU instance hardware and reduced their pricing in the past, so the cloud-based solution may compare even more favorably in the future. 

Another potential advantage of running GPU-based simulations in cloud computing environments is the opportunity for parallelism with no up-front costs. A single MuMax3 simulation fully occupies a GPU during execution, so that the number of simulations that can be run simultaneously is limited to the number of available GPUs. Purchasing and maintaining an array of GPUs is unlikely to prove cost effective in typical use cases, especially since simulation workloads are often sporadic. On the other hand, each researcher using AWS can launch 5 parallel instances on-demand or up to 20 instances on a reservation basis\cite{AWS}. This easily-scaled computing capacity, which has largely motivated the interest in cloud-computing platforms for web services, can offer significant speed-ups when running large batches of simulations without incurring anything beyond the standard hourly instance charges. 

\section{MuCloud Software}

\begin{figure}
\centering
\includegraphics[width=0.85\columnwidth]{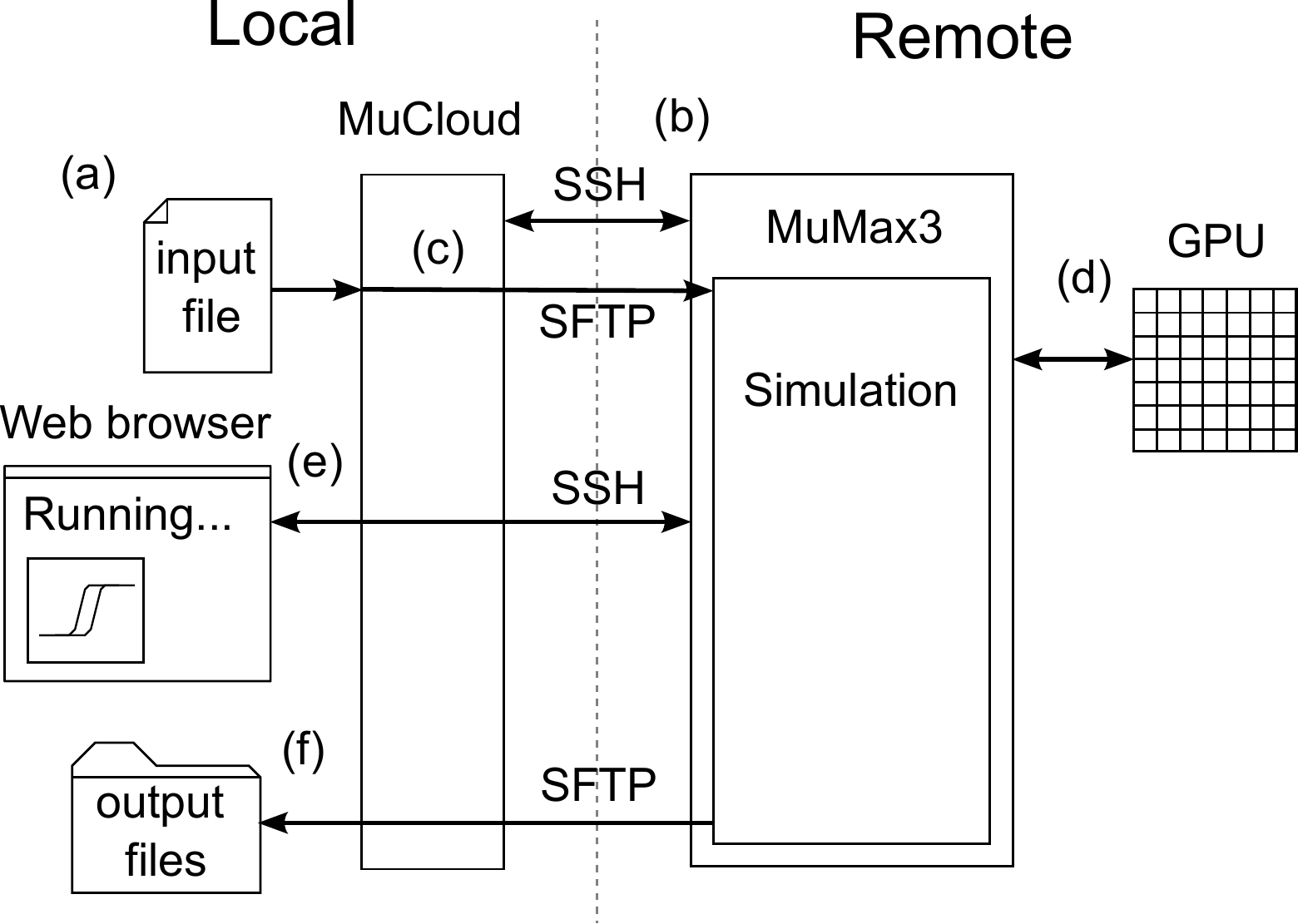}
\caption{\label{mucloud}Illustration of MuCloud use. (a) On the local computer, a normal MuMax3 input file is passed into MuCloud, which (b) connects to a new or existing AWS GPU instance. (c) The input file is transferred to the remote instance and MuMax3 is started. (d) The simulation runs on the remote GPU, (e) during which time the web browser interface of MuMax3 is accessible on the local computer. (f) Upon completion, output files generated by MuMax3 are transferred back to the local computer by MuCloud. }
\end{figure}

We have developed an open-source Python script, MuCloud, that runs MuMax3 simulations on AWS GPU instances irrespective of the user's local operating system. This code can be obtained on GitHub\cite{MuCloud} along with full documentation regarding its installation, capabilities, and use. The operation of the script is detailed in Figure \ref{mucloud}. For security purposes, Secure Shell (SSH) and Secure File Transfer Protocol (SFTP) ensure that all data is encrypted while passing between the local and remote computers. The MuMax3 web interface is accessible so that the local user can control and monitor the simulations on the remote instance in real time.

\section{Performance}

We quantify the performance of MuMax3 on AWS GPU instances by timing the execution of a magnetic-field-driven reversal simulation for a variety of simulated sample sizes. We compare to the performance of a CPU-based solver, OOMMF\cite{Donahue1999} (running on 4 threads), and determine the system size regime where the GPU-based cloud-computing approach has superior performance. A simple magnetic-field-driven reversal problem is chosen for the performance benchmark because micromagnetic packages tend to diverge in their implementation of more complex phenomena (\textit{e.g.} the inclusion of spin-transfer torque). No attempts are made to optimize the execution in either simulator, and therefore we expect that this comparison is accurate for typical-use cases. Our analysis differs from those presented by Arne \textit{et al.}\cite{Vansteenkiste2014} and Lopez-Diaz \textit{et al.}\cite{Lopez-Diaz2012} in that we consider the total simulation execution time, instead of the solver step time.

\begin{figure}
\centering
\includegraphics[width=0.85\columnwidth]{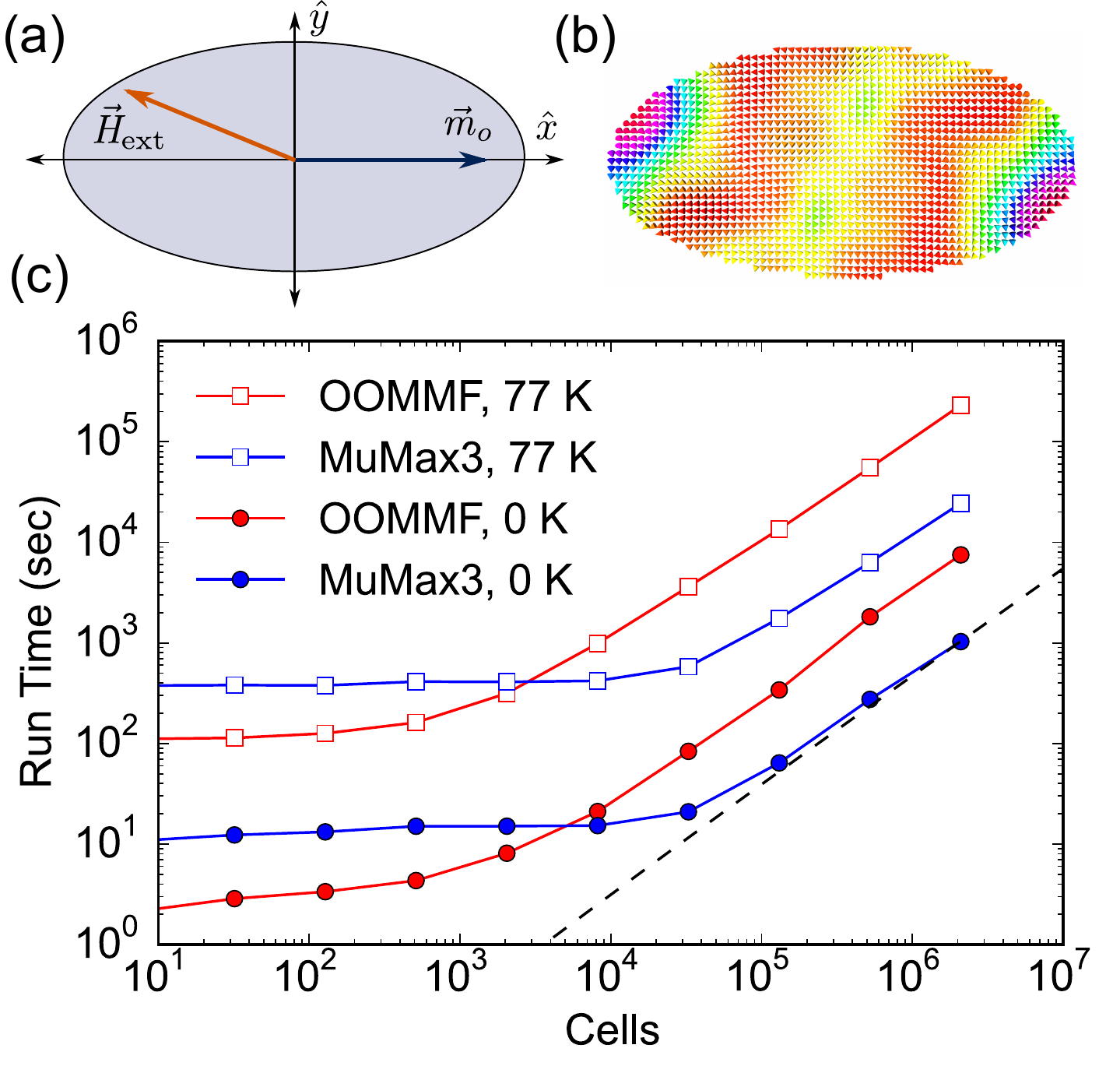}
\caption{\label{run-time}(a) Geometry of the field-reversal problem for comparing solvers. (b) Micromagnetic texture during field-reversal. (c) Run time required to simulate 1 ns of the problem using either (blue) MuMax3 or (red) OOMMF, as a function of the number of simulation cells ($2 N \times N \times 1$). (solid circles) Zero-temperature simulations are performed with Dormand-Prince (RK45) evolution, while (open squares) finite temperature (77 K) simulations use the Heun method. (black dashed line) Simulations approach $\mathcal{O}(N\log{}N)$ complexity at high cell count.}
\end{figure} 

Our benchmark problem examines an elliptical thin-film Permalloy (NiFe) nano-magnet as illustrated in Figure \ref{run-time}(a). We use an exchange interaction strength of $A_{ex} = 13 \times 10^{-12}$ J/m, saturation magnetization of $M_s = 800 \times 10^3$ A/m, and an initial magnetization ($\vec{m}_o$) saturated in the $+\hat{x}$ direction. The magnet has an aspect ratio of 2, with sizes ranging from $16\,\times\,8\,\times\,2$ nm$^3$ to $4000\,\times\,2000\,\times\,2$ nm$^3$.  The simulation region is discretized into $2 N \times N \times 1$ cubes of 8 nm$^3$ volume. $N$ is restricted to multiples of 8 to avoid performance penalties in the fast-Fourier transform (FFT) algorithm that arise from non-``seven smooth" system dimensions.

We test the zero- and finite-temperature performance using the Dormand-Prince (RK45) and Heun methods, respectively, for evolving the Landau-Liftshitz equation. We apply an external magnetic field of $\vec{H}_{\mathrm{ext}} = -60 \hat{x} + 20 \hat{y}$ mT with a step-like time dependence, which causes the sample magnetization to reverse by domain nucleation and propagation following smaller amplitude precessional pre-switching oscillations. Figure \ref{run-time}(b) illustrates the micromagnetic texture of the magnetization during the reversal process.

\begin{figure}
\centering
\includegraphics[width=0.85\columnwidth]{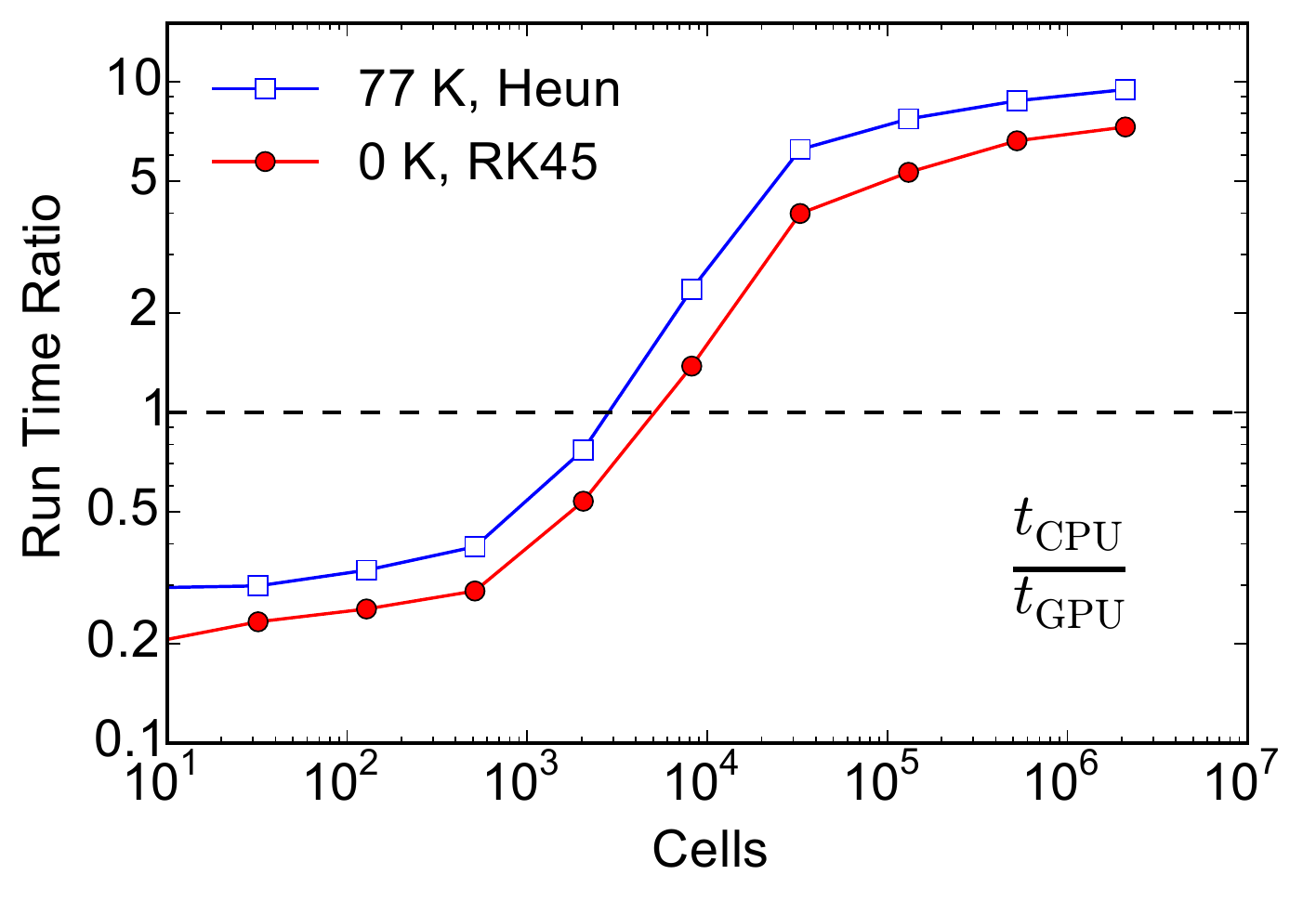}
\caption{\label{speed-up}The ratio of the CPU and GPU run times associated with simulating the problem with either (red) zero-temperature with Dormand-Prince (RK45) or (blue) finite-temperature (77 K) with the Heun method. Above approximately 5,000 cells, the large parallelism of MuMax3's GPU based solvers provides a performance advantage compared to the CPU-based simulation.}
\end{figure}

Figure \ref{run-time}(c) illustrates the execution time required to simulate 1 ns of the field-reversal problem. The additional burden of computing thermal fields at finite temperatures raises the execution time by almost two orders of magnitude, but does not significantly change the qualitative dependence on the system size. Figure \ref{speed-up} shows the relative ratio between the CPU and GPU run times. Below 5,000 cells, the CPU simulations have better performance. In our problem, this corresponds to a magnet with dimensions at or below $128 \,\times\, 64 \,\times\, 2$ nm$^3$ ($N = 64$). In this regime the reduced speed of the GPU clock compared to the CPU, the delays for memory transfers to and from GPU memory, and other execution latencies limit the performance of the graphics card since its instrinsic parallelism is not taken advantage of. In larger simulations, however, the GPU-based simulations provide a speed-up factor from 7 to 10. Both the CPU and GPU calculations exhibit similar $\mathcal{O}(N\log{}N)$ complexity for large sizes, which is expected from the FFT operations involved in calculating the demagnetization field. This causes the speed-up factor associated with using the GPU to saturate for large systems.

Although current micromagnetic simulation packages are hard-coded to utilize either CPU or GPU resources, it should be possible to utilize a heterogeneous computing approach to enable seamless and intelligent transitions between processor types based on the geometry and characteristics of the simulated system. While CPU bound simulation codes remain the better choice for conducting small simulations, the nearly order-of-magnitude reduction in simulation time for large systems constitutes a significant advantage to using GPU-accelerated micromagnetics on cloud computing services.

\section{Conclusion}

Cloud computing services provide a means for researchers to obtain the performance enhancements of GPU-based micromagnetic simulations without investing in specialized computer hardware. This opens new possibilities such as simultaneous simulations across a large number of remote instances. We present an open-source program (MuCloud) that allows MuMax3 simulations to be run on AWS instances, so that researchers can easily access this new avenue for micromagnetics. With these tools, we demonstrate that a nearly ten-fold performance enhancement can be obtained over CPU-based micromagnetic codes when simulating large systems.

We acknowledge Barry Robinson and Jim Entwood for introducing us to Amazon Web Services.  This research was supported by the NSF (DMR-1010768) and IARPA (W911NF-14-C-0089).


\bibliography{references}

\begin{thebibliography}{10}
\expandafter\ifx\csname url\endcsname\relax
  \def\url#1{\texttt{#1}}\fi
\expandafter\ifx\csname urlprefix\endcsname\relax\def\urlprefix{URL }\fi
\expandafter\ifx\csname href\endcsname\relax
  \def\href#1#2{#2} \def\path#1{#1}\fi

\bibitem{Fidler2000}
J.~Fidler, T.~Schrefl, {Micromagnetic modelling - the current state of the
  art}, Journal of Physics D: Applied Physics 33 (2000) R135--R156.
\newblock \href {http://dx.doi.org/10.1088/0022-3727/33/15/201}
  {\path{doi:10.1088/0022-3727/33/15/201}}.

\bibitem{Berkov2008}
D.~V. Berkov, J.~Miltat, {Spin-torque driven magnetization dynamics:
  Micromagnetic modeling}, Journal of Magnetism and Magnetic Materials 320
  (2008) 1238--1259.
\newblock \href {http://dx.doi.org/10.1016/j.jmmm.2007.12.023}
  {\path{doi:10.1016/j.jmmm.2007.12.023}}.

\bibitem{Thiaville2012}
A.~Thiaville, S.~Rohart, E.~Jue, V.~Cros, A.~Fert, {Dynamics of Dzyaloshinskii
  domain walls in ultrathin magnetic films}, Europhysics Letters 100~(5) (2012)
  57002.
\newblock \href {http://dx.doi.org/10.1209/0295-5075/100/57002}
  {\path{doi:10.1209/0295-5075/100/57002}}.

\bibitem{Perez2014}
N.~Perez, E.~Martinez, L.~Torres, S.~Woo,
  \href{http://arxiv.org/abs/1401.3526}{{Chiral magnetization textures
  stabilized by the Dzyaloshinskii-Moriya interaction during spin-orbit torque
  switching}}, Applied Physics Letters 104~(May) (2014) 092403.
\newline\urlprefix\url{http://arxiv.org/abs/1401.3526}

\bibitem{Vansteenkiste2013}
A.~Vansteenkiste, B.~{Van De Wiele}, L.~Dupr\'{e}, B.~{Van Waeyenberge}, D.~{De
  Zutter}, {Implementation of a finite-difference micromagnetic model on GPU
  hardware}, International Journal of Numerical Modelling: Electronic Networks,
  Devices and Fields 26~(March 2012) (2013) 366--375.
\newblock \href {http://dx.doi.org/10.1002/jnm.1835}
  {\path{doi:10.1002/jnm.1835}}.

\bibitem{Vansteenkiste2014}
A.~Vansteenkiste, J.~Leliaert, M.~Dvornik, M.~Helsen, F.~Garcia-Sanchez,
  B.~{Van Waeyenberge}, {The design and verification of MuMax3}, AIP Advances
  4~(10) (2014) 107133.
\newblock \href {http://dx.doi.org/10.1063/1.4899186}
  {\path{doi:10.1063/1.4899186}}.

\bibitem{Vansteenkiste2011}
A.~Vansteenkiste, B.~{Van de Wiele}, {MuMax: A new high-performance
  micromagnetic simulation tool}, Journal of Magnetism and Magnetic Materials
  323~(21) (2011) 2585--2591.
\newblock \href {http://dx.doi.org/10.1016/j.jmmm.2011.05.037}
  {\path{doi:10.1016/j.jmmm.2011.05.037}}.

\bibitem{Chang2011}
R.~Chang, S.~Li, M.~V. Lubarda, B.~Livshitz, V.~Lomakin, {FastMag: Fast
  micromagnetic simulator for complex magnetic structures (invited)}, Journal
  of Applied Physics 109~(7) (2011) 1--6.
\newblock \href {http://dx.doi.org/10.1063/1.3563081}
  {\path{doi:10.1063/1.3563081}}.

\bibitem{MicroMagnum}
A.~Drews, G.~Selke, C.~Abert, T.~Gerhardt, J.~M. Meyer, C.~Darsow-Fromm, B.~K.
  {M. Menzel}, D.~P.~F. M\"{o}ller, {MicroMagnum},
  \url{http://micromagnum.informatik.uni-hamburg.de/} (2011).

\bibitem{Lopez-Diaz2012}
L.~Lopez-Diaz, D.~Aurelio, L.~Torres, E.~Martinez, M.~a. Hernandez-Lopez,
  J.~Gomez, O.~Alejos, M.~Carpentieri, G.~Finocchio, G.~Consolo, {Micromagnetic
  simulations using Graphics Processing Units}, Journal of Physics D: Applied
  Physics 45 (2012) 323001.
\newblock \href {http://dx.doi.org/10.1088/0022-3727/45/32/323001}
  {\path{doi:10.1088/0022-3727/45/32/323001}}.

\bibitem{Vliet2011}
J.~van Vliet, F.~Paganelli, {Programming Amazon EC2}, O'Reilly Media, 2011.

\bibitem{AWS}
{Amazon Web Services, Inc.}, \url{http://aws.amazon.com}.

\bibitem{MuCloud}
C.~L. Jermain, G.~E. Rowlands, {MuCloud},
  \url{http://ralph-group.github.io/mucloud} (2014--2015).

\bibitem{Donahue1999}
M.~Donahue, D.~Porter, {OOMMF User's Guide, Version 1.0}, Tech. rep., National
  Institute of Standards and Technology, Gaithersburg, MD (1999).

\end{thebibliography}

\end{document}